\documentclass[
reprint,
superscriptaddress,
amsmath,
amssymb,
aps,
]{revtex4-2}

\usepackage{array}
\usepackage{changes}
\usepackage{tabularx}
\usepackage{makecell}
\usepackage{graphicx}
\usepackage{dcolumn}
\usepackage{bm}
\usepackage{float}
\usepackage{placeins}
\usepackage{soul}
\usepackage{xcolor}
\usepackage[
    colorlinks=true,
    citecolor=blue,
    linkcolor=blue,
    urlcolor=blue
]{hyperref}

\begin{document}

\title{Magnetic-Order-Driven Nonlinear Photocurrents in $\mathcal{PT}$-Broken Antiferromagnetic MnS$_2$ and Altermagnetic MnSe$_2$
}

\author{Fu Li}
\affiliation{Institute of Materials Science, Technical University of Darmstadt,
Darmstadt 64287, Germany}

\author{Rajyavardhan Ray}
\affiliation{Department of Physics, Birla Institute of Technology Mesra,
Ranchi 835215, Jharkhand, India}

\author{Xiaoxiong Liu}
\affiliation{International Center for Quantum Materials,Beijing Institute of Technology, Zhuhai, 519000, China}

\author{Babu Baijnath Prasad}
\affiliation{Institute for Solid State Physics, The University of Tokyo,
Kashiwa, Chiba 277-8581, Japan}

\author{Chen Shen}
\affiliation{Suzhou Laboratory, Suzhou 215021, China}

\author{Yaqian Guo}
\affiliation{Leibniz Institute for Solid State and Materials Research Dresden, 01069 Dresden, Germany}

\author{Jeroen van den Brink}
\affiliation{Leibniz Institute for Solid State and Materials Research Dresden, 01069 Dresden, Germany}
\affiliation{W{\"u}rzburg-Dresden Cluster of Excellence ctd.qmat, 01069 Dresden, Germany}
\affiliation{Institute of Theoretical Physics, Technische Universit{\"a}t Dresden, 01062 Dresden, Germany}

\author{Hongbin Zhang}
\email{hongbin.zhang@tu-darmstadt.de}
\affiliation{Institute of Materials Science, Technical University of Darmstadt,
Darmstadt 64287, Germany}

\author{Harish K. Singh}
\email{harish.singh@plus.ac.at}
\affiliation{Department of Chemistry and Physics of Materials,
University of Salzburg, Jakob-Haringer-Strasse 2a,
5020 Salzburg, Austria}


\begin{abstract}
Nonlinear photocurrent responses generally require inversion-symmetry breaking, but its origin can be magnetic rather than structural. In this study, we investigate the bulk photovoltaic effect in antiferromagnetic MnS$_2$ and altermagnetic MnSe$_2$, two pyrite-type semiconductors whose magnetic ordering breaks inversion symmetry $\mathcal{P}$ and its combination with time reversal $\mathcal{PT}$, while their ionic lattices remain centrosymmetric. We calculate the shift current (SC) and injection current (IC) responses under linearly and circularly polarized light with and without spin--orbit coupling (SOC) and identify their symmetry-allowed tensor components using the corresponding magnetic and spin point groups, respectively. Linear SC and circular IC are symmetry-allowed in both MnS$_2$ and MnSe$_2$ irrespective of SOC, whereas linear IC and circular SC remain forbidden in MnS$_2$ but become allowed in MnSe$_2$ only when SOC is included. These results highlight the importance of SOC and the corresponding MPG analysis for a complete description of nonlinear charge photocurrents in magnetic materials. Beyond charge responses, MnSe$_2$ supports linear shift and circular injection spin photocurrents in the nonrelativistic limit, whereas all spin-photocurrent responses are forbidden in MnS$_2$. Our findings therefore establish inversion symmetry breaking induced by compensated magnetic order as a route to generating and controlling nonlinear photocurrents. 
\end{abstract}
\maketitle
\underline{\it Introduction} Nonlinear optical responses provide a powerful route to probe quantum materials because they are highly sensitive to crystal symmetry, magnetic order, and the quantum geometry of electronic states~\cite{wang2019ferroicity,gutzler2021light,weiner2017light}. Among these responses, second-order direct photocurrents are macroscopic manifestations of the bulk photovoltaic effect (BPVE) and reflect inversion $\mathcal{P}$ symmetry breaking, providing insight into noncentrosymmetric electronic structures. Under uniform illumination, the BPVE generates a direct current in a homogeneous material without a $p$--$n$ junction, offering photovoltaic functionality beyond the Shockley--Queisser paradigm~\cite{dai2023recent,zhao2026recent}. The two principal intrinsic mechanisms are shift current (SC) and injection current (IC), with the former arising from an inversion-asymmetric displacement of the electronic charge center and the latter from an asymmetric carrier velocity distribution during optical excitation~\cite{kim2017shift,panday2019injection,liu2020circular}.

A necessary condition for shift and injection photocurrents is the breaking of $\mathcal{P}$ symmetry. In nonmagnetic noncentrosymmetric materials, $\mathcal{P}$ symmetry is broken, while time-reversal symmetry $\mathcal{T}$ is preserved. This allows time-even charge responses permitted by the crystal point-group symmetries such as linear SC and circular IC, driven by linearly and circularly polarized light, respectively~\cite{avdoshkin2025multistate,de2017quantized}. Such responses have been widely explored in structurally noncentrosymmetric systems, in a variety of materials including ferroelectric oxides, hybrid perovskites, two-dimensional materials, topological materials, chiral crystals, and semiconductors~\cite{rangel2017large,li2026medium,yang2009photovoltaic,tan2016shift,sauer2023shift,osterhoudt2019colossal,tan2016enhancement,ishii2025giant,dey2024correlation,wang2026strain,wei2026shift}.

\FloatBarrier
\begin{table*}[!t]
\caption{Summary of symmetry-allowed second-order charge photocurrent tensor components for MnS$_2$ and MnSe$_2$ with and without SOC, obtained from the MPG and  SPG, respectively. Here, $\sigma_{ijk}$ and $\eta_{ijk}$ denote the shift and injection current response tensors, respectively. In the Jahn symbols, $V$ represents a polar vector and $aV$ denotes a time-reversal-odd (MPG) or spin-space-odd polar vector (SPG)~\cite{lihm2022comprehensive}.}
\label{tab:unified_charge_photocurrent_components}

\centering
\renewcommand{\arraystretch}{1.20}
\setlength{\tabcolsep}{4pt}

\begin{tabular}{ccccccc}
\hline\hline
Channel
& \makecell{Part\\}
& \makecell{Jahn\\symbol}
& \multicolumn{2}{c}{With SOC: MPG}
& \multicolumn{2}{c}{Without SOC: SPG} \\
\cline{4-5} \cline{6-7}
&
&
& MnS$_2$   
& MnSe$_2$  
& MnS$_2$ 
& MnSe$_2$ \\
\hline

\makecell{Linear SC\\}
& \makecell{Real\\symmetric}
& $V[V^2]$
& \makecell[l]{
$\sigma_{xxx}$, $\sigma_{xyy}$, $\sigma_{xzz}$,\\
$\sigma_{yxy}=\sigma_{yyx}$, \\
$\sigma_{zxz}=\sigma_{zzx}$
}
& \makecell[l]{
$\sigma_{zxx}$, $\sigma_{zyy}$,$\sigma_{zzz}$,\\
$\sigma_{xzx}=\sigma_{xxz}$, \\
$\sigma_{yyz}=\sigma_{yzy}$
}
& \makecell[l]{
$\sigma_{xxx}$, $\sigma_{xyy}$, $\sigma_{xzz}$,\\
$\sigma_{yxy}=\sigma_{yyx}$, \\
$\sigma_{zxz}=\sigma_{zzx}$
}
& \makecell[l]{
$\sigma_{zxx}$, $\sigma_{zyy}$,$\sigma_{zzz}$,\\
$\sigma_{xzx}=\sigma_{xxz}$, \\
$\sigma_{yyz}=\sigma_{yzy}$
} \\

\hline

\makecell{Linear IC\\}
& \makecell{Real\\symmetric}
& $aV[V^2]$
& 0
& \makecell[l]{
$\eta_{xxx}$, $\eta_{xyy}$, $\eta_{xzz}$,\\
$\eta_{yxy}=\eta_{yyx}$,\\ $\eta_{zxz}=\eta_{zzx}$\\
}
& 0
& 0 \\

\hline

\makecell{Circular IC\\}
& \makecell{Imaginary\\antisymmetric}
& $V\{V^2\}$
& \makecell[l]{
$\eta_{yxy}=-\eta_{yyx}$,\\
$\eta_{zzx}=-\eta_{zxz}$
}
& \makecell[l]{
$\eta_{xzx}=-\eta_{xxz}$,\\
$\eta_{yyz}=-\eta_{yzy}$
}
& \makecell[l]{
$\eta_{yxy}=-\eta_{yyx}$,\\
$\eta_{zzx}=-\eta_{zxz}$
}
& \makecell[l]{
$\eta_{xzx}=-\eta_{xxz}$,\\
$\eta_{yyz}=-\eta_{yzy}$
} \\

\hline

\makecell{Circular SC\\}
& \makecell{Imaginary\\antisymmetric}
& $aV\{V^2\}$
& 0
& \makecell[l]{
$\sigma_{yxy}=-\sigma_{yyx}$,\\
$\sigma_{zzx}=-\sigma_{zxz}$
}
& 0
& 0 \\

\hline\hline
\end{tabular}
\end{table*}

Magnetic order offers an alternative route to $\mathcal{P}$-symmetry breaking even when the ionic lattice remains centrosymmetric. In several layered and low-dimensional antiferromagnets, it breaks $\mathcal{P}$ and $\mathcal{T}$ individually while preserving their product $\mathcal{PT}$, as reported for CrI$_3$~\cite{zhang2019switchable,gudelli2020antiferromagnetism,fei2021pt}, MnBi$_2$Te$_4$~\cite{fei2020giant,wang2020electrically}, Ti$_4$C$_3$~\cite{sufyan2026evidence}, and EuSn$_2$As$_2$~\cite{pi2023magnetic}. For charge photocurrents, $\mathcal{PT}$ symmetry suppresses the time-even linear SC and circular IC but may allow their time-odd counterparts, linear IC and circular SC, depending on the magnetic point group (MPG). 

The symmetry-enforced suppression of certain charge-photocurrent channels raises the question of whether their spin counterparts can survive. Since charge and spin photocurrents obey distinct symmetry selection rules, a spin photocurrent may remain finite even when the corresponding charge response vanishes~\cite{xu2021pure,yang2026symmetry,jiang2025nonlinear}. In $\mathcal{PT}$-symmetric AFMs, opposite-spin contributions may add in the spin channel while canceling in the charge channel, permitting spin linear SC and circular IC~\cite{fei2021pt,xu2021pure}. Such spin currents can be generated on ultrafast timescales by optical excitation without an applied voltage~\cite{sato2016laser}. However, if $\mathcal{P}$ remains an exact symmetry of the full magnetic or spin group, all intrinsic second-order charge and spin photocurrents vanish, requiring a higher-order mechanism such as the bias-assisted third-order response proposed for centrosymmetric altermagnets~\cite{dong2026band}. These contrasting regimes motivate the study of second-order charge and spin photocurrents in systems where magnetic order breaks both $\mathcal{P}$ and $\mathcal{PT}$.

In this context, MnS$_2$ and MnSe$_2$ provide representative bulk magnetic semiconductors: the former hosts a collinear AFM state that preserves the antiunitary translation symmetry $\mathcal{T}\mathbf{t}$, whereas the latter is an altermagnet (ALM)~\cite{jaeschke2024supercell}. In both compounds, long-range magnetic order develops below their respective Néel temperatures ($T_N\sim50~\mathrm{K}$)~\cite{kimber2015spin,hastings1959antiferromagnetic,itoh1977magnetic,chattopadhyay1987magnetic} and breaks both $\mathcal{P}$ and $\mathcal{PT}$ in the bulk, while the atomic positions remain centrosymmetric. Therefore, the $\mathcal{P}$-symmetry breaking relevant to nonlinear photocurrents is magnetic and electronic in origin rather than driven by a polar lattice distortion~\cite{matsuda2025multiferroic,kocsis2026frustration}. This distinguishes MnS$_2$ and MnSe$_2$ from conventional multiferroics, where $\mathcal{P}$-symmetry breaking is commonly associated with ionic displacements.

Motivated by this distinction, we investigate second-order photocurrents in pyrite-type MnS$_2$ and MnSe$_2$. We calculate all four charge-photocurrent channels, namely linear SC, circular SC, linear IC, and circular IC, with and without SOC, and evaluate their spin counterparts only without SOC because the spin-resolved channels remain well defined in this limit. The allowed tensor components, determined by MPGs when SOC is included and by spin point groups (SPGs) when it is neglected, are summarized for MnS$_2$ and MnSe$_2$ in Tables~\ref{tab:unified_charge_photocurrent_components} and~\ref{tab:unified_spin_photocurrent_components_spg}. The importance of this distinction was recently demonstrated in altermagnetic $\alpha$-Fe$_2$O$_3$, where the experimentally observed XMCD, a linear magneto-optical response, was shown to require SOC and to be governed by the MPG rather than by spin-group symmetry~\cite{mallon2026revealing}. An analogous distinction emerges for nonlinear photocurrents in MnSe$_2$. Specifically, linear IC and circular SC vanish without SOC and emerge only when SOC is included. Since SOC is intrinsic to real materials, an SPG analysis alone cannot fully describe the MnSe$_2$ response, and an MPG analysis is therefore required to capture all four charge-photocurrent channels. Beyond these symmetry changes, SOC has only a minor effect on the MnS$_2$ spectra but modifies specific spectral features in MnSe$_2$. To identify the microscopic origins of these responses, we analyze momentum-resolved photocurrents, optical conductivity, and the joint density of states (JDOS). This combined approach clarifies how magnetic symmetry and electronic structure shape nonlinear charge and spin photocurrents and how SOC modifies the charge-photocurrent responses of MnS$_2$ and MnSe$_2$.

\underline{\it Methods}
Under coherent optical excitation, a direct photocurrent can arise from second-order light--matter interactions. Within the electric-dipole approximation, the shift current contribution is described by
\begin{equation}
J_i^{\mathrm{Shift}}=\sigma_{ijk}(\omega)E_j(\omega)E_k(-\omega),
\end{equation}
whereas the injection current contribution is given by
\begin{equation}
J_i^{\mathrm{Injection}}=\eta_{ijk}(\omega)E_j(\omega)E_k(-\omega).
\end{equation}
Here, $E_j$ and $E_k$ denote the optical electric-field components along the $j$ and $k$ directions, respectively, and the index $i$ specifies the direction of the direct current. Further details of the photocurrent formalism are provided in Section~S1 of the Supporting Information (SI).

\begin{figure}[!htp]
	\centering
	\includegraphics[width=1.0\linewidth]{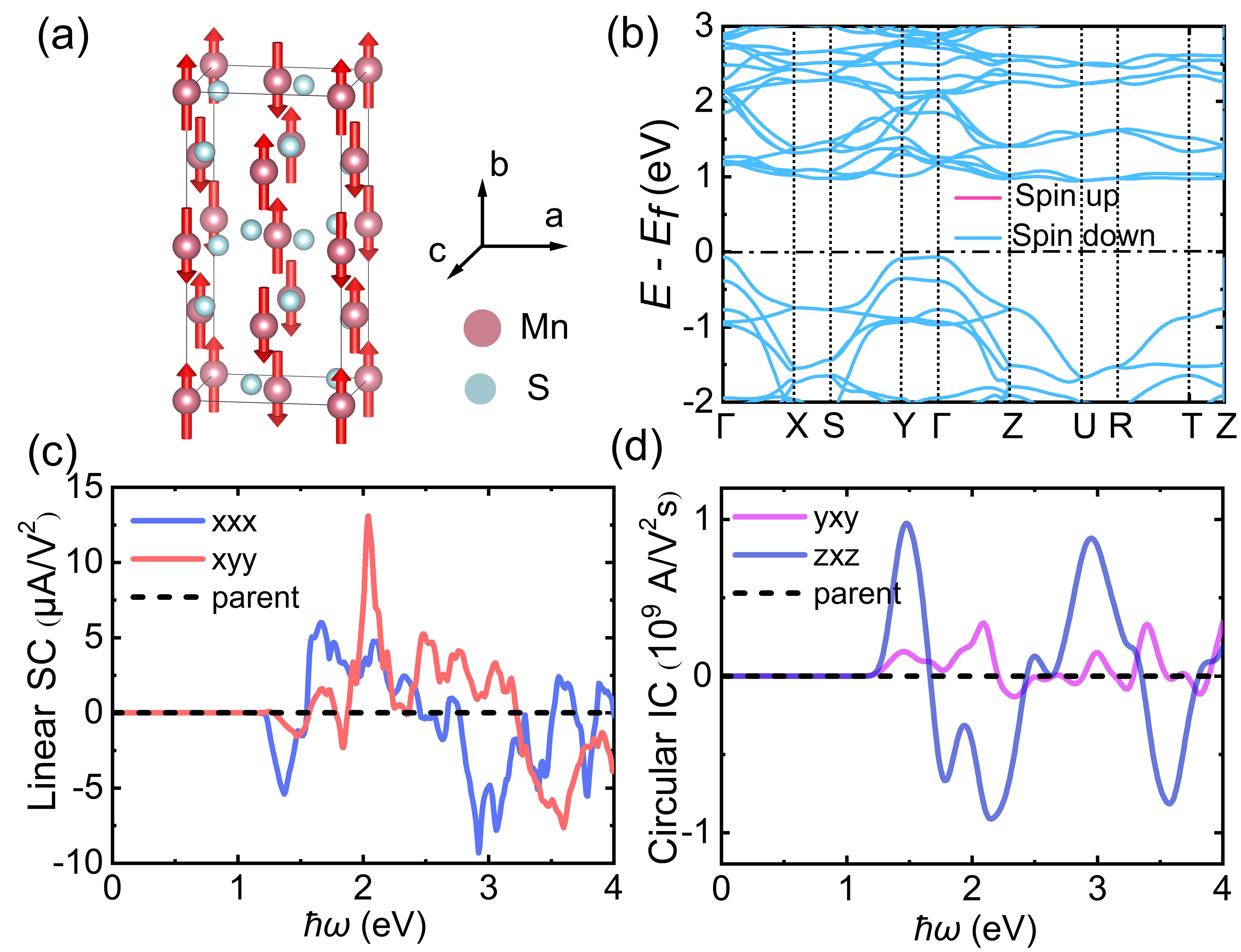}
	\caption{\label{fig:MnS2-crystal-structure}Crystal and electronic structures and charge photocurrent responses of AFM MnS$_2$ with SOC. Dashed line (parent) corresponds to the centrosymmetric, nonmagnetic cubic phase of MnS$_2$. (a) Magnetic crystal structure. (b) Calculated electronic band structure without SOC. (c) Linear SC spectra for two dominant components; (d) Circular IC spectra for two dominant components.  All symmetry-allowed charege photocurrent components are shown in \textcolor{blue}{Fig.~S7}.}
\end{figure}

Density functional theory (DFT) calculations were performed using the projector-augmented-wave (PAW) method~\cite{blochl1994,kresse1999}, as implemented in the Vienna \textit{Ab initio} Simulation Package (VASP)~\cite{kresse1993abinitio,kresse1994abinitio,kresse1996efficient,kresse1996efficiency}. The exchange-correlation interactions were described by the Perdew--Burke--Ernzerhof (PBE) implementation of the generalized gradient approximation~\cite{perdew1996generalized}. The Wannier tight-binding Hamiltonians were constructed using Wannier90~\cite{mostofi2014updated}, and the linear and circular
SC and IC tensors were evaluated with
WannierBerri~\cite{tsirkin2021high}. Computational details are provided in Section~S1 of the SI.

\underline{\it Crystal and magnetic structures} 
MnS$_2$ and MnSe$_2$ crystallize in the centrosymmetric cubic pyrite structure in the paramagnetic phase, with space group $Pa\bar{3}$ (No.~205)~\cite{kimber2015spin,hastings1959antiferromagnetic,wang2006facile,itoh1977magnetic,chattopadhyay1987magnetic}. The Mn$^{2+}$ ions span a fcc sublattice and are octahedrally coordinated by S$_2^{2-}$ or Se$_2^{2-}$ dimers (Figs.~\ref{fig:MnS2-crystal-structure}(a) and \textcolor{blue}{~S1-S2}). Below their respective Néel temperatures
($T_N\sim50~\mathrm{K}$), both compounds develop collinear magnetic order with commensurate propagation vectors $\mathbf{k}=(0,\frac{1}{2},0)$ and $\mathbf{k}=(\frac{1}{3},0,0)$, respectively, and are classified as supercell magnets. The corresponding magnetic and spin-group symmetries are summarized in \textcolor{blue}{Tables~S1--S3}.

Supercell magnets are often associated with type-IV magnetic space groups, as in MnS$_2$, which belongs to $P_bca2_1$ and exhibits spin-degenerate bands (Fig.~\ref{fig:MnS2-crystal-structure}(b)). In AFM MnS$_2$, the opposite-spin sublattices are connected by the antiunitary $\mathcal{T}\mathbf{t}$ operation (\textcolor{blue}{Table~S2 and Fig.~S1}). However, not all supercell magnets belong to type-IV magnetic space groups. For instance, MnSe$_2$ is a supercell altermagnet described by the type-III magnetic space group $Pca^{\prime}2_1^{\prime}$~\cite{jaeschke2024supercell}. In MnSe$_2$, the opposite-spin sublattices are related by a mirror operation rather than by $\mathcal{PT}$ or $\mathcal{T}\mathbf{t}$ symmetry, yielding momentum-dependent altermagnetic spin splitting (Figs. \textcolor{blue}{S2, S4, and Table~S3}). 

\begin{figure}[!htp]
\centering
\includegraphics[width=1.0\linewidth]{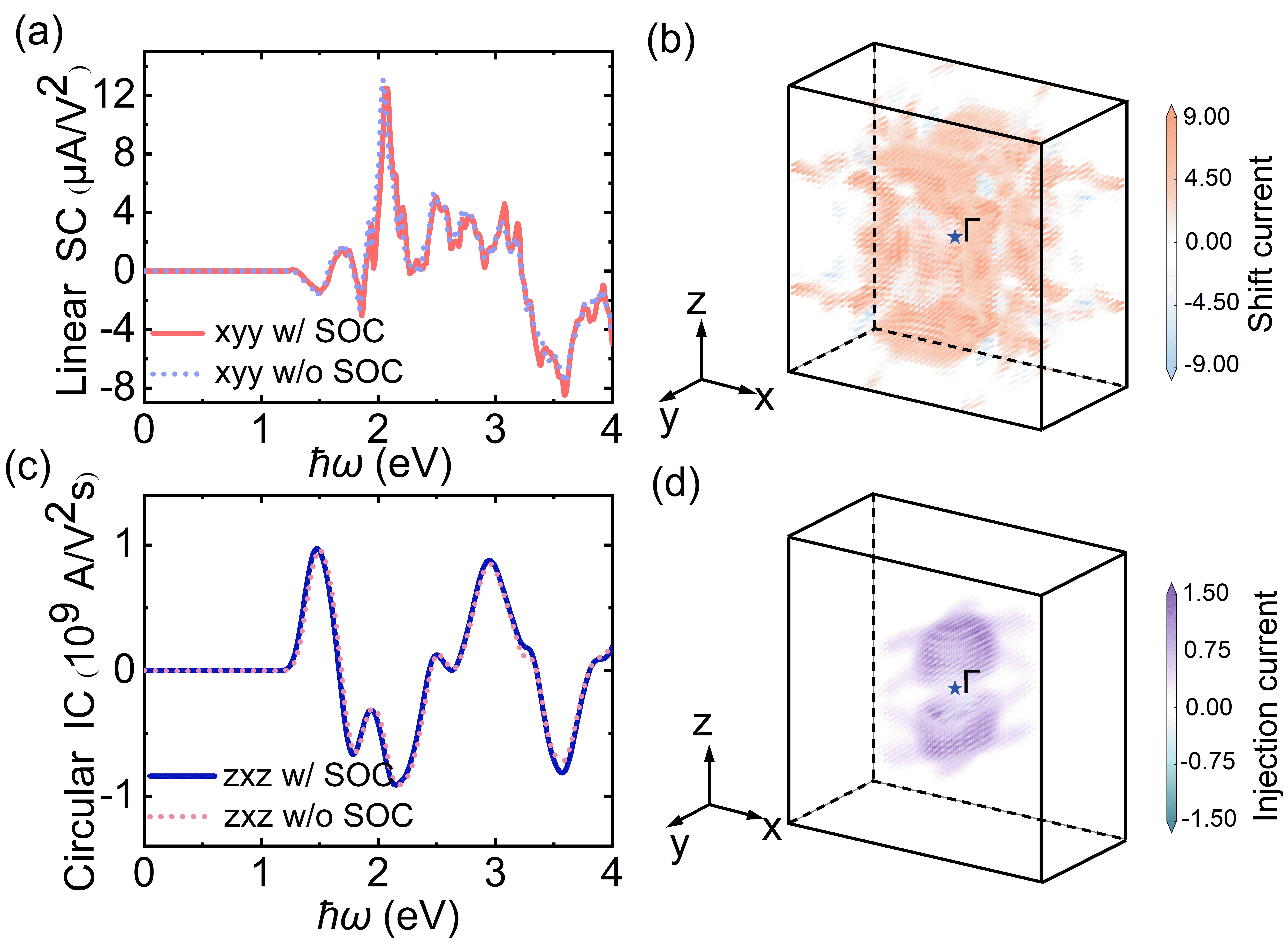}
\caption{Comparison of the charge photocurrent responses of AFM MnS$_2$ calculated with and without SOC. (a) Linear SC spectra. (b) Brillouin-zone distribution of the $\sigma_{xyy}$ linear SC component at $\hbar\omega=2.04~\mathrm{eV}$ with SOC. (c) Circular IC spectra. (d) Brillouin-zone distribution of the $\eta_{zxz}$ circular IC component at $\hbar\omega=1.48~\mathrm{eV}$ with SOC.}
\label{fig:mns2_bpve}
\end{figure}

Despite their distinct magnetic symmetries, both compounds exhibit frustrated AFM orders that break inversion symmetry, as detailed in \textcolor{blue}{Section~S2}. In MnS$_2$, for example, the inversion-related sites Mn2($\downarrow$) and Mn6($\uparrow$) carry opposite spins (\textcolor{blue}{Fig.~S1}). Because spin remains unchanged under spatial inversion, $\mathcal{P}$ maps each moment onto a site occupied by the opposite moment and, therefore, does not preserve the magnetic configuration. The corresponding polar magnetic point groups are $\mathrm{mm2}.1^{\prime}$ for MnS$_2$ and $\mathrm{m^{\prime}m2^{\prime}}$ for MnSe$_2$. Because their ionic lattices remain centrosymmetric, the induced polarizations are purely electronic, reaching $567~\mu\mathrm{C}/\mathrm{m}^2$ at U= 4.0 eV in MnS$_2$~\cite{kocsis2026frustration}. 

\underline{\it Non-linear Photocurrents} 
The agreement between the calculated tensor components in \textcolor{blue}{Figs.~S6--S7 and S10--S12} and the symmetry-allowed and forbidden components in Tables~\ref{tab:unified_charge_photocurrent_components} and~\ref{tab:unified_spin_photocurrent_components_spg} confirms our symmetry analysis, whose underlying theory and transformation rules are detailed in Section~S3 of the SI (\textcolor{blue}{Tables~S4--S27}). With SOC, MnS$_2$ belongs to the grey MPG $\mathrm{mm2}.1^{\prime}=G+1^{\prime}G$, where $G=\mathrm{mm2}$ and $1^{\prime}$ denotes pure time reversal. The time-even linear SC and circular IC are symmetry-allowed with five and two finite tensor components, respectively. In contrast, because linear IC and circular SC are time-odd, $1^{\prime}$ reverses all their tensor components, forcing both tensors to vanish (\textcolor{blue}{Tables S4-S7}). The symmetry-allowed photocurrent tensor components are the same with and without SOC. In the latter case, spin-space inversion $-1$ in the SPG operations $[-1\parallel G]$ imposes the same sign reversal (\textcolor{blue}{Tables~S12--S15}).

\begin{figure}[!htp]
	\centering
	\includegraphics[width=1.0\linewidth]{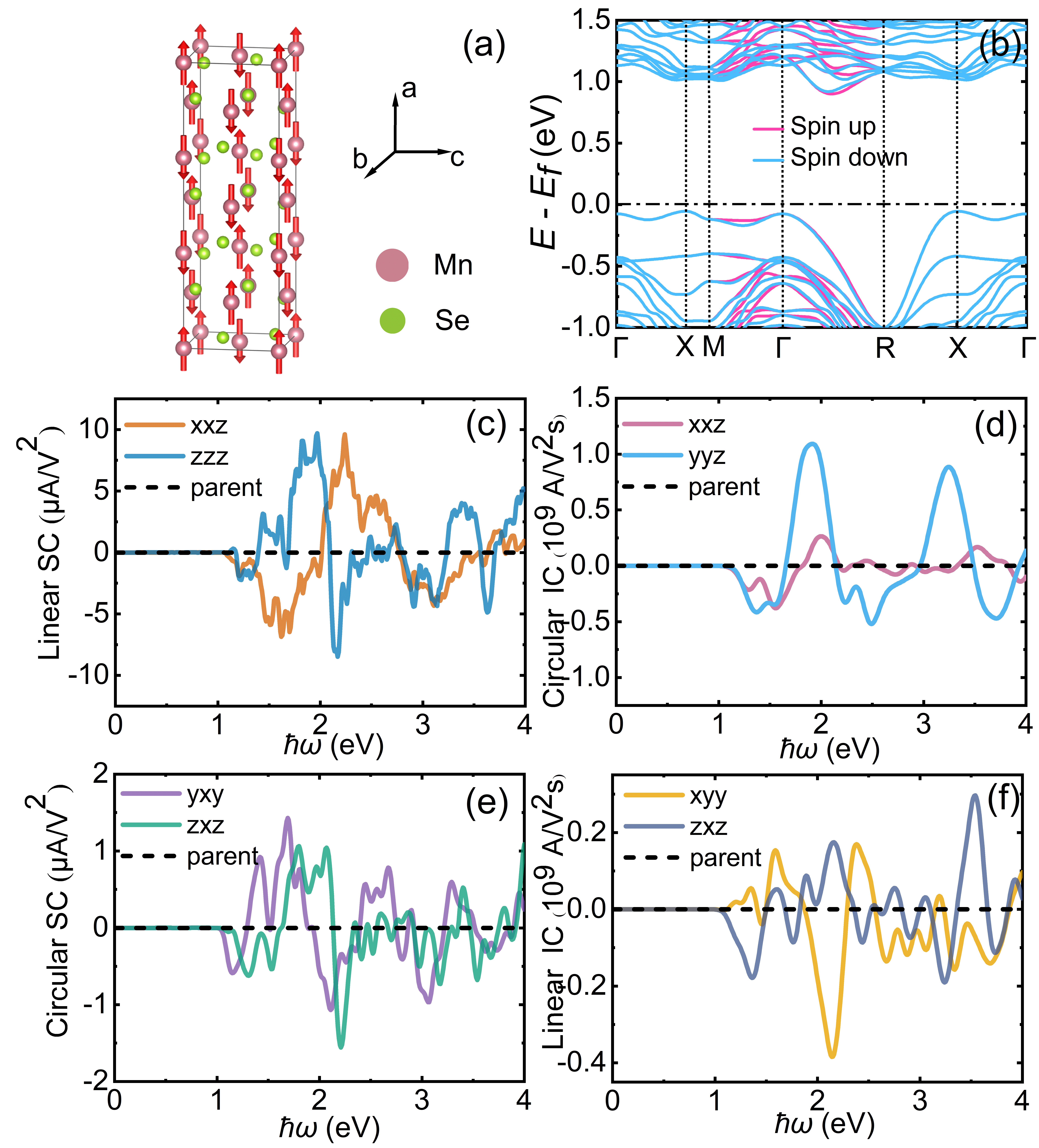}
\caption{Crystal and electronic structures and charge photocurrent responses of ALM MnSe$_2$ with SOC. Dashed line (parent) corresponds to the centrosymmetric, nonmagnetic cubic phase. (a) Magnetic crystal structure. (b) Calculated electronic band structure without SOC. (c) Linear SC spectra for the two dominant components. (d) Circular IC spectra. (e) Circular SC spectra. (f) Linear IC spectra for the two dominant components. All symmetry-allowed charge photocurrent components of MnSe$_2$ are shown in \textcolor{blue}{Figs.~S11 and S12}.}
\label{fig:mnse2_structure_band}
\end{figure}

We next examine how these symmetry-allowed responses manifest in the calculated linear SC and circular IC spectra of MnS$_2$, shown in Figs.~\ref{fig:MnS2-crystal-structure}(c) and (d), respectively. The photocurrent onset near $1.2$~eV, slightly above the indirect band gap of $0.98$~eV (Fig.~\ref{fig:MnS2-crystal-structure}(b)), coincides with the onset of optically allowed interband transitions (\textcolor{blue}{Fig. S16}). At $\hbar\omega=2.04~\mathrm{eV}$, the linear SC response $\sigma_{xyy}$ component reaches approximately $13.09~\mu\mathrm{A}/\mathrm{V}^2$, comparable to the value reported for well-known bulk AFM BiFeO$_3$~\cite{young2013prediction,prasad2024magnetism}. This linear SC peak lies close to a pronounced maximum in the $y$-polarized optical conductivity $(\sigma^S_{yy})$, whereas the JDOS shows no corresponding maximum (\textcolor{blue}{Fig. S16}). The suppression of $\sigma_{xxx}$ and $\sigma_{xzz}$ relative to $\sigma_{xyy}$ in this energy window is consistent with the pronounced optical anisotropy of the magnetic state ($\mathbf{k}=(0,\frac{1}{2},0)$, $\mathbf{S}\parallel y$), also evident in the linear optical conductivity (\textcolor{blue}{Fig. S16}). In comparison, the circular IC response is dominated by the $\eta_{zxz}$ component throughout the photon-energy range (Fig.~\ref{fig:MnS2-crystal-structure}(d)). It reaches a pronounced peak of $0.98\times10^9\mathrm{A}/(\mathrm{V}^2\mathrm{s})$ near $1.47$~eV, suggesting significant contributions from transitions involving states near the band edges.

Having established that SOC does not change the symmetry-allowed charge photocurrent tensor components in MnS$_2$, we further examine its effect on the magnitude and momentum-space distribution of the allowed responses. We focus on two representative tensor components: the dominant linear SC component $\sigma_{xyy}$ and the dominant circular IC component $\eta_{zxz}$. Their spectra with and without SOC are shown in Figs.~\ref{fig:mns2_bpve}(a) and \ref{fig:mns2_bpve}(c), respectively. In both cases, the spectra calculated with and without SOC are nearly identical, indicating that SOC has only a minor effect on the overall photocurrent response. To further clarify their microscopic origins, we analyze the $k$-resolved distributions at the corresponding peak photon energies, $2.04$ and $1.48$~eV. As shown in Figs.~\ref{fig:mns2_bpve}(b) and \ref{fig:mns2_bpve}(d), the linear SC component $\sigma_{xyy}$ receives contributions from extended regions of the Brillouin zone (BZ), whereas the circular IC component $\eta_{zxz}$ is dominated by localized hot spots. The two-dimensional projections further clarify these distinct momentum-space origins (\textcolor{blue}{Figs.~S8 and S9}). These results indicate that the linear SC and circular IC are governed primarily by magnetic-order-induced inversion-symmetry breaking and the underlying nonrelativistic band geometry, rather than by SOC-induced band splitting.

We next turn to altermagnetic MnSe$_2$, which belongs to the MPG $\mathrm{m^{\prime}m2^{\prime}}$ and exhibits photocurrent selection rules distinct from those of MnS$_2$ (Table~\ref{tab:unified_charge_photocurrent_components}). Linear SC and circular IC retain five and two finite tensor components, respectively, both with and without SOC. In contrast, linear IC and circular SC are forbidden without SOC but acquire five and two finite components, respectively, when SOC is included (\textcolor{blue}{Table S8-S11 and S16-S19}). Thus, unlike MnS$_2$, MnSe$_2$ exhibits SOC-dependent time-odd linear IC and circular SC responses, whereas the time-even linear SC and circular IC responses remain symmetry-allowed in both materials irrespective of SOC.

\FloatBarrier
\begin{table}[!t]
\caption{Summary of symmetry-allowed second-order \emph{spin}
photocurrent tensor components without SOC,
obtained using the SPG. The spin-current response
tensors are denoted by $\sigma^s_{i;jk}$ and $\eta^s_{i;jk}$, where
$s$ is the spin-polarization direction and $i$ is the current
direction. In the generalized Jahn symbols, $M$ denotes the independent
spin-polarization vector; the prefactor $a$ indicates spin-space determinant factor under an SPG operation.}
\label{tab:unified_spin_photocurrent_components_spg}

\centering
\renewcommand{\arraystretch}{1.20}
\setlength{\tabcolsep}{5pt}

\begin{tabular}{cccc}
\hline\hline
Channel
& \makecell{Generalized\\Jahn\\symbol}
& \multicolumn{2}{c}{Without SOC: SPG} \\
\cline{3-4}
&
& MnS$_2$
& MnSe$_2$ \\
\hline

\makecell{Linear SC}
& $MV[V^2]$
& $0$
& \makecell[l]{
$\sigma^x_{x;yz}=\sigma^x_{x;zy}$,\\
$\sigma^x_{y;xz}=\sigma^x_{y;zx}$,\\
$\sigma^x_{z;xy}=\sigma^x_{z;yx}$
} \\

\hline

\makecell{Linear IC}
& $aMV[V^2]$
& $0$
& $0$ \\

\hline

\makecell{Circular IC}
& $MV\{V^2\}$
& $0$
& \makecell[l]{
$\eta^x_{x;yz}=-\eta^x_{x;zy}$,\\
$\eta^x_{y;xz}=-\eta^x_{y;zx}$,\\
$\eta^x_{z;xy}=-\eta^x_{z;yx}$
} \\

\hline

\makecell{Circular SC}
& $aMV\{V^2\}$
& $0$
& $0$ \\

\hline\hline
\end{tabular}
\end{table}

\begin{figure*}[!htp]
\centering
\includegraphics[width=0.7\linewidth]{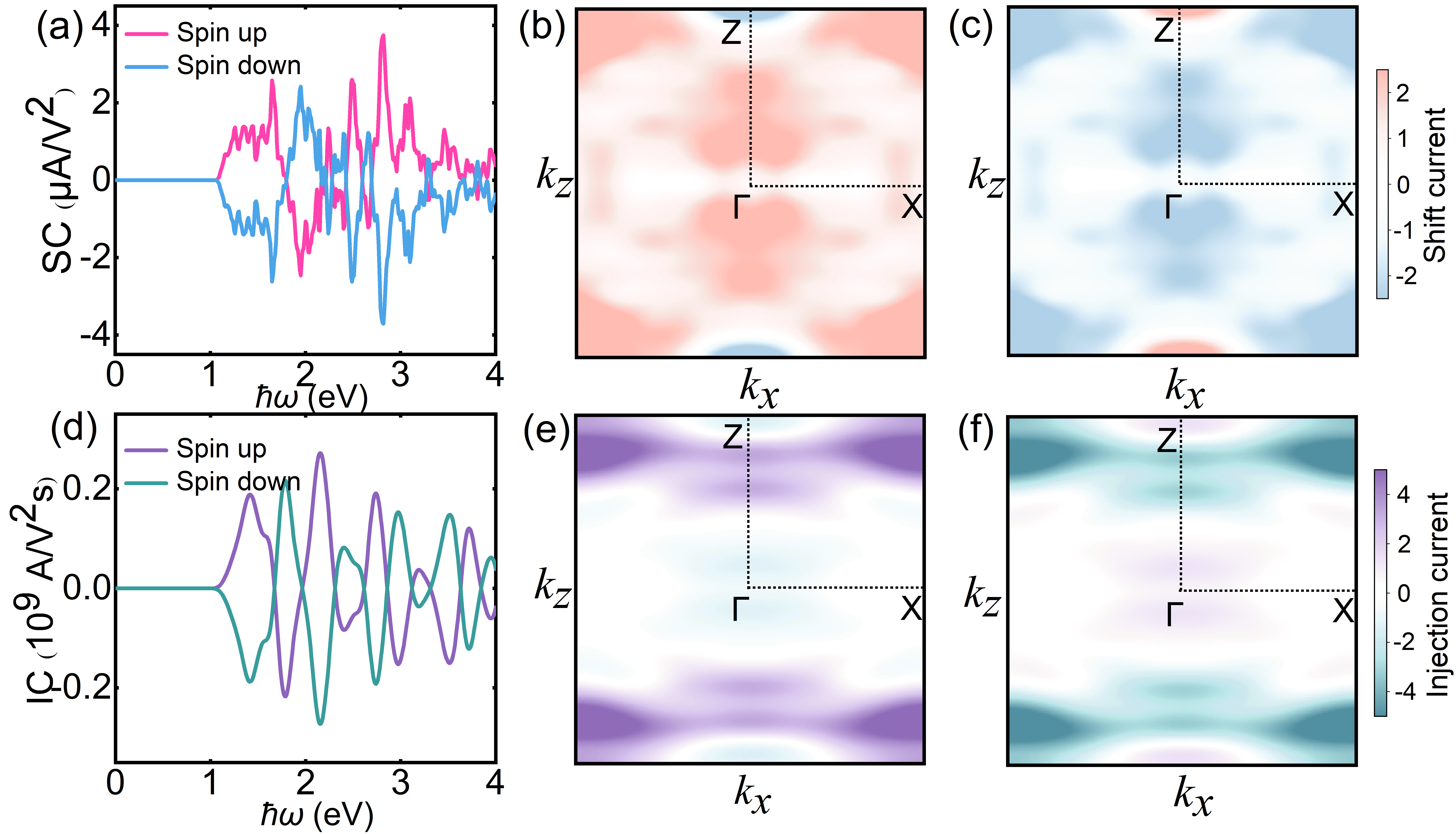}
\caption{Spin-resolved photocurrent responses of MnSe$_2$ without SOC. (a) Linear SC component $\sigma^x_{z;xy}$. (b,c) Corresponding $k$-resolved distributions at $\hbar\omega = 2.82~\text{eV}$, projected onto the $k_x$--$k_z$ plane for spin-up and spin-down, respectively. (d) Circular IC component $\eta^x_{z;xy}$. (e,f) Corresponding $k$-resolved distributions at $\hbar\omega = 2.15~\text{eV}$ for spin-up and spin-down, respectively. Dashed lines indicate the $\Gamma$--$X$ and $\Gamma$--$Z$ directions.}
\label{fig:mnse2_spin_current}
\end{figure*}

To determine the magnitudes and spectral characteristics of these responses, Figure~\ref{fig:mnse2_structure_band} shows that the centrosymmetric
parent phase of MnSe$_2$ exhibits no second-order photocurrent response,
whereas magnetic ordering breaks $\mathcal{P}$ symmetry and activates
all four photocurrent channels. The linear SC reaches peak magnitudes of $\sim10~\mu\mathrm{A}/\mathrm{V}^2$ between $1.5$ and $2.5$~eV, only marginally smaller than those in MnS$_2$, whereas the circular SC is considerably weaker, with a peak of $\sim1.5~\mu\mathrm{A}/\mathrm{V}^2$ in the same energy range (Fig.~\ref{fig:mnse2_structure_band} (c) and (e)). Interestingly, although the magnetic ordering gives rise to a strongly anisotropic linear optical conductivity response with a dominant xx component (\textcolor{blue}{Fig. S17}), this anisotropy is not reflected in the SC spectra above 1.5 eV, where different current directions exhibit comparable peak magnitudes, highlighting the important role of the corresponding shift-vector contributions. Thus, large shift current responses in MnSe$_2$ are obtained in a much wider photon energy range compared to MnS$_2$.

The linear SC component $\sigma_{zzz}$ reaches a large peak of $9.70~\mu\mathrm{A}/\mathrm{V}^2$ at $\hbar\omega=1.97~\mathrm{eV}$ [Fig.~\ref{fig:mnse2_structure_band}(c)], originating from extended positive contributions across the BZ that are partially offset by smaller localized negative contributions (\textcolor{blue}{Fig.~S13}). The circular IC reaches $\sim1\times10^9~\mathrm{A}/(\mathrm{V}^2\mathrm{s})$, comparable to MnS$_2$, whereas the linear IC is considerably weaker, peaking at $\sim0.4\times10^9~\mathrm{A}/(\mathrm{V}^2\mathrm{s})$ [Figs.~\ref{fig:mnse2_structure_band}(d) and (f)]. At $\hbar\omega=1.92~\mathrm{eV}$, the circular IC component $\eta_{yyz}$ likewise receives broadly distributed contributions across the BZ (\textcolor{blue}{Fig.~S14}). Comparison of the spectra calculated with and without SOC reveals more pronounced changes in specific features of MnSe$_2$ than in MnS$_2$ (\textcolor{blue}{Fig.~S15}). Near $2.5~\mathrm{eV}$, the prominent $\sigma_{zzz}$ peak without SOC disappears when SOC is included, while the corresponding feature in $\eta_{yyz}$ is reduced in magnitude.

Beyond charge photocurrents, we analyze spin photocurrents without SOC using SPG symmetry, as summarized in Table~\ref{tab:unified_spin_photocurrent_components_spg}, with the transformations of tensor components detailed in \textcolor{blue}{Tables~S20--S27}. All four spin-photocurrent channels are symmetry-forbidden in MnS$_2$. By contrast, MnSe$_2$ allows linear SC and circular IC, each with three independent $s_x$-polarized tensor components, while linear IC and circular SC remain forbidden. For each allowed spin-photocurrent component, the mirror-related spin
sectors generate equal and opposite charge-photocurrent contributions.
Therefore, the corresponding charge tensor component vanishes, whereas the spin response remains finite, yielding a component-resolved pure spin current. Other symmetry-allowed charge-photocurrent components of MnSe$_2$, including transverse ones, remain finite, although their corresponding spin-photocurrent components vanish.

The spin-resolved spectra of the symmetry-allowed linear SC and circular IC in ALM MnSe$_2$ are shown in Fig.~\ref{fig:mnse2_spin_current} (a) and (d). For both responses, the spin-up and spin-down channels contribute with opposite signs. At the selected photon energies, $\hbar\omega=2.82~\mathrm{eV}$ for linear SC and $2.15~\mathrm{eV}$ for circular IC, the respective spin-resolved responses reach approximately $3.74~\mu\mathrm{A}/\mathrm{V}^2$ and $0.28\times10^9~\mathrm{A}/(\mathrm{V}^2\mathrm{s})$, producing particularly strong spin-channel contrast. The corresponding $k$-resolved maps (Fig.~\ref{fig:mnse2_spin_current}(b,c) and (e,f)) show sign-reversed patterns between the two spin channels. The linear SC weight is broadly distributed across the projected $k_x$--$k_z$ plane, whereas the circular IC weight is concentrated near the projected BZ boundary. These spectral and momentum-space differences reflect spin-dependent interband processes, with linear SC governed by optical transition weights and shift vectors and circular IC by helicity-dependent optical transition probabilities and carrier-velocity differences. Together, these features demonstrate that nonrelativistic ALM spin splitting supports finite spin photocurrents in both the linear SC and circular IC channels.

\underline{\it Conclusions \& Outlook}
Our results demonstrate that a centrosymmetric ionic lattice does not preclude a sizable bulk photovoltaic response when compensated magnetic order breaks electronic inversion symmetry, a mechanism recently shown to induce electronic polarization~\cite{kocsis2026frustration} and identified here as the origin of the finite photocurrents. The calculated photocurrent tensor components agree with our symmetry analysis. Linear SC and circular IC are symmetry-allowed in both MnS$_2$ and MnSe$_2$ with and without SOC, unlike in $\mathcal{PT}$-symmetric antiferromagnets, where both time-even charge responses are symmetry-forbidden. Moreover, SOC activates the linear IC and circular SC charge responses in MnSe$_2$, both of which vanish in its absence. This demonstrates that SOC and the corresponding MPG analysis are required to capture all four charge-photocurrent channels in MnSe$_2$. Without SOC, MnSe$_2$ supports component-resolved pure spin photocurrents through the linear-SC and circular-IC channels. The momentum-resolved distributions further show that different responses originate from distinct regions of the BZ, demonstrating that symmetry determines which channels are allowed, while the electronic structure controls their strength and spectral character. This distinction between responses allowed without SOC and those enabled by SOC provides a symmetry-based design principle for magnetic nonlinear optoelectronics.

\underline{\it Acknowledgments}  
The authors would like to thank Dr. Oleg Janson and Dr. Hao Wang for valuable discussions and insightful suggestions. The authors acknowledge the computing time provided on the Lichtenberg high-performance computer by the NHR Center NHR4CES at TU Darmstadt. Fu Li acknowledges support from the China Scholarship Council. H.~Zhang also acknowledges support from the Deutsche Forschungsgemeinschaft (DFG, German Research Foundation), Project-ID 463184206. Harish K. Singh acknowledges the EuroHPC Joint Undertaking for granting access to the VEGA supercomputer at IZUM, Slovenia, through project EHPC-REG-2025R01-197-EuroHPC.

\nocite{*}
\bibliography{reference}

\end{document}